\documentclass[sigplan,nonacm]{acmart}

\usepackage[ruled,lined]{algorithm2e}
\usepackage{xspace}
\usepackage{subfig}
\usepackage[dvipsnames]{xcolor}
\usepackage{tikz}
\usepackage{flushend}
\usepackage{soul}

\AtBeginDocument{%
  }

\setcopyright{acmlicensed}
\copyrightyear{2018}
\acmYear{2018}
\acmDOI{XXXXXXX.XXXXXXX}
\acmConference[Conference acronym 'XX]{Make sure to enter the correct
  conference title from your rights confirmation email}{June 03--05,
  2018}{Woodstock, NY}
\acmISBN{978-1-4503-XXXX-X/2018/06}

\graphicspath{{figs/}}

\ifx\figurename\undefined \def\figurename{Figure}\fi
\renewcommand{\figurename}{Fig.}
\newcommand{\para}[1]{\textit{\textbf{#1}} }
\renewcommand{\subparagraph}[1]{\underline{\textit{#1}} }

\newcommand{\Sect}[1]{Sec.~\ref{#1}}
\newcommand{\Fig}[1]{Fig.~\ref{#1}}

\newcommand{\mode}[1]{\underline{\textsc{#1}}\xspace}

\newcommand{\proj}{\textsc{Nebula}\xspace}

\definecolor{myorange}{RGB}{255, 212, 121}
\newcommand{\circnum}[1]{%
  \tikz[baseline=(char.base)]{
    \node[
      circle,
      draw=black,
      line width=2pt,       % Outer boundary thickness
      fill=black,
      inner sep=3pt,
      minimum size=12pt
    ] (outer) {};

    \node[
      circle,
      draw=black,
      line width=0.8pt,     % Inner border
      fill=orange!50,
      inner sep=2.2pt,
      minimum size=10pt
    ] (char) {\sffamily\bfseries #1};
  }%
}

\renewcommand{\hl}[1]{{#1}}

\begin{document}

\title[\proj: Enable City-Scale 3D Gaussian Splatting in Virtual Reality]{\proj: Enable City-Scale 3D Gaussian Splatting in Virtual Reality via Collaborative Rendering and Accelerated Stereo Rasterization}

\author[He Zhu et al.]{He Zhu$^{1}$ \quad Zheng Liu$^{1}$ \quad Xingyang Li$^{1}$ \quad Anbang Wu$^{1}$ \quad Jieru Zhao$^{1}$ \quad Fangxin Liu$^{1}$ \\ Yiming Gan$^{3}$  \quad Jingwen Leng$^{1,2,\dagger}$ \quad Yu Feng$^{1,2,\dagger}$}

\affiliation{%
  \institution{$^{1}$Shanghai Jiao Tong University \quad 
               $^{2}$Shanghai Qi Zhi Institute \\ 
    $^{3}$Institute of Computing Technology, Chinese Academy of Sciences}
  \country{}
}

\email{{zhcon16, distilledw, brucelee_sjtu, anbang, zhao-jieru, liufangxin, leng-jw, y-feng}@sjtu.edu.cn}
\email{ganyiming@ict.ac.cn}

\thanks{$\dagger$Corresponding authors}

% \author{He Zhu}
% \author{Zheng Liu}
% \author{Xingyang Li}
% \author{Anbang Wu}
% \author{Jieru Zhao}
% \author{Fangxin Liu}
% \author{Yiming Gan}
% \author{Jingwen Leng\thanks{Corresponding authors}}
% \author{Yu Feng\footnotemark[1]}

% \affiliation{%
%   \institution{$^{1}$Shanghai Jiao Tong University \quad
%                $^{2}$Shanghai Qi Zhi Institute \quad
%                $^{3}$Institute of Computing Technology, Chinese Academy of Sciences}
%   \country{}
% }

% % grouped emails (SJTU group)
% \email{{zhcon16, distilledw, brucelee_sjtu, anbang, zhao-jieru, liufangxin, leng-jw, y-feng}@sjtu.edu.cn}

% % single email
% \email{ganyiming@ict.ac.cn}

\begin{abstract}

3D Gaussian splatting (3DGS) has drawn significant attention in the architectural community recently. 
However, current architectural designs often overlook the 3DGS scalability, making them fragile for extremely large-scale 3DGS.
Meanwhile, the VR bandwidth requirement makes it impossible to deliver high-fidelity and smooth VR content from the cloud.

We present \proj, a coherent acceleration framework for large-scale 3DGS collaborative rendering. 
Instead of streaming videos, \proj streams intermediate results after the LoD search, reducing 1925\% data communication between the cloud and the client.
To further enhance the motion-to-photon experience, we introduce a temporal-aware LoD search in the cloud that tames the irregular memory access and reduces redundant data access by exploiting temporal coherence across frames.
On the client side, we propose a novel stereo rasterization that enables two eyes to share most computations during the stereo rendering with bit-accurate quality.
With minimal hardware augmentations, \proj achieves 2.7$\times$ motion-to-photon speedup and reduces 1925\% bandwidth over lossy video streaming.

\end{abstract}


  
% %% A "teaser" image appears between the author and affiliation
% %% information and the body of the document, and typically spans the


\maketitle

\section{Introduction}
\label{sec:intro}

\begin{figure*}[t]
    \centering
    \includegraphics[width=0.95\textwidth]{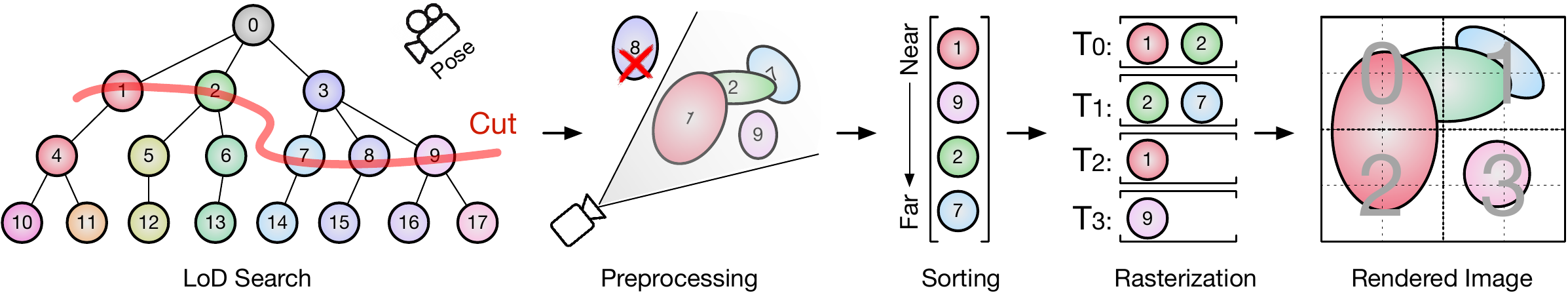}
    \caption{The rendering pipeline for large-scale 3DGS consists of four stages: LoD search, preprocessing, sorting, and rasterization.
    First, LoD search traverses the LoD tree to determine a set of Gaussians with a desired LoD granularity.
    The result Gaussians form a ``cut'' that separates the top and bottom of the LOD tree.
    Then, the Gaussians on the cut go through a sequence of operations, i.e., preprocessing, sorting, and rasterization, to render an image, similar to the small-scale 3DGS pipelines~\cite{kerbl20233d}.}
    \label{fig:pipeline}
\end{figure*}

Neural rendering is ushering in a renaissance in computer graphics by enabling photorealistic and view-dependent rendering, with much higher speeds than conventional ray tracing~\cite{pharr2023physically, deng2017toward, pantaleoni2010hlbvh}.
In recent years, neural rendering has drawn significant attention in the architectural community~\cite{feng2025lumina, ye2025gaussian, feng2024potamoi, lee2024gscore, li2025uni, lee2025vr, lin2025metasapiens, durvasula2025arc, he2025gsarch, lee2023neurex, rao2022icarus, li2023instant, mubarik2023hardware, li2022rt, fu2023gen, feng2024cicero, song2024srender, liu2025cambricon}, with 3D Gaussian splatting (3DGS) standing out due to its compact representation and superior rendering performance.

While prior 3DGS accelerator designs~\cite{feng2025lumina, ye2025gaussian, feng2024potamoi, lee2024gscore, li2025uni, lee2025vr, lin2025metasapiens, durvasula2025arc, he2025gsarch} achieve real-time mobile rendering for small-scale scenes~\cite{barron2022mipnerf360, hedman2018deep, Knapitsch2017}, they often overlook the scalability challenge of 3DGS, making their designs fragile for large-scale rendering (e.g., city-scale)~\cite{kerbl2024hierarchical, ren2024octree, liu2024citygaussian, li2023matrixcity, wu2025blockgaussian}.
As shown in \Sect{sec:ch:local}, the memory requirement for such scenes can reach up to 66 GB, far exceeding the typical memory capacity ($<$12~GB) of devices in virtual reality (VR)~\cite{questprospec, htcvivespec, visionprospec}. 
This memory gap motivates us to design a collaborative rendering framework that leverages the resources available in the cloud to overcome the memory constraints of VR devices.
% \fixme{Only motivate the memory overhead, however, we also need to motivate the compute overhead from VR headset.}

Despite numerous solutions for remote or collaborative rendering~\cite{meng2020coterie, xie2021q, leng2019energy, zhao2020deja, wen2023post0, xu2023edge, zhao2021holoar}, they primarily target conventional video streaming. 
Such approaches paired with the HEVC codecs~\cite{sullivan2012overview, h264, h265} often require over 1~Gbps for 4K VR content at 90~FPS~\cite{mangiante2017vr}.
To alleviate the bandwidth pressure, we make two key insights unique to large-scale 3DGS:
1) in virtual 3D scenes, the number of newly visible Gaussians introduced by continuous pose changes remains roughly constant; and
2) the memory requirement of large-scale 3DGS peaks during the initial level-of-detail (LoD) search, but drops sharply in the subsequent stages.

By leveraging these insights, we propose \proj, a collaborative rendering framework tailored for 3DGS at \textit{infinite scale}.
Rather than streaming fully rendered images, \proj transmits the intermediate results after the initial LoD search, i.e., Gaussians required for subsequent stages, to the client.
We show that \proj requires a 1925\% lower bandwidth compared to conventional video streaming.
Additionally, \proj also exhibits strong scalability: its bandwidth demand is less susceptible to resolution or frame rate increases.

Algorithmically, \proj makes three core contributions.
First, we propose a \textit{temporal-aware LoD search} algorithm in \Sect{sec:algo:search} that can be deployed upon existing GPUs out-of-the-box.
Specifically, our algorithm regulates the DRAM access by streamingly processing data in LoD search and leverages the temporal similarity across frames to avoid unnecessary data accesses.
Second, we design a \textit{runtime Gaussian management} system in \Sect{sec:algo:mem} that compresses and transmits only the non-overlapped Gaussians across the adjacent frames to further reduce the data transfer between the cloud and the client.
Third, for VR rendering, where two tightly paired stereo displays need to be rendered, we introduce a novel \textit{stereo rasterization} pipeline in \Sect{sec:algo:stereo} that exploits \textit{triagulation}~\cite{hartley2003multiple, szeliski2010computer}, a widely-used technique in computer vision to share most of the computations in the remaining pipeline while still producing bit-accurate images.
% skip the repetitive stages, i.e., preprocessing and sorting, in subsequent pipeline, while further reducing the computation of rasterization, one key bottleneck in 3DGS.

Architecturally, we show that our stereo rasterization can be easily integrated into any mainstream 3DGS accelerators with minimal hardware augmentations (\Sect{sec:arch}).
Overall, \proj achieves 2.7$\times$ motion-to-photon speedup and reduces bandwidth by 1925\% compared to video streaming.
On the cloud side, our temporal-aware LoD search delivers up to 52.7$\times$ speedup compared to off-the-shelf GPU implementations.
On the client side, our stereo rasterization achieves up to 21.7$\times$ speedup and 5.3$\times$ speedup, compared to a mobile Ampere GPU and the state-of-the-art accelerators~\cite{lee2024gscore, ye2025gaussian}, respectively, all with minimal hardware overhead.

The contributions of this paper are as follows:
\begin{itemize}
    \item A collaborative rendering framework that is tailored for large-scale 3DGS with great scalability.
    \item Two techniques, temporal-aware LoD search and stereo rasterization, that accelerate both sides of computation by up to 52.7$\times$ and 21.7$\times$.
    \item Our architecture achieves 2.6$\times$ speedup and reduces bandwidth by 1925\% over lossy video streaming.
    % all with minimal hardware augmentation.
\end{itemize}

\section{Background}
\label{sec:bg}

In this section, we first give a brief background on remote rendering in \Sect{sec:bg:remote}. Then, we introduce the general rendering pipeline for large-scale 3DGS in \Sect{sec:bg:pipeline}.

\subsection{Remote Rendering}
\label{sec:bg:remote}

\para{Video Streaming.}
The mainstream approach to remote rendering in AR/VR is video streaming.
A client first transmits a pose to a remote server, which then renders an image corresponding to that pose and streams it back to the client.
Lastly, the client displays the received image on the screen. 

To reduce the bandwidth requirements, video streaming typically applies various video compression techniques, which can be classified into two main categories: conventional methods~\cite{h264, h265} and DNN-based methods~\cite{deng2021deep, liu2019dsic, wodlinger2022sasic}. While DNN-based compressions offer high compression rates, they are often too compute-intensive for latency-sensitive applications such as VR/AR. 
Nowadays, most real-time video streaming systems for immersive applications still rely on conventional compression techniques, which strike a balance between compression efficiency and latency.

\para{Collaborative Rendering.}
Recent studies leverage collaborative rendering, which harnesses the compute power of both remote servers and local client devices~\cite{xie2021q, xu2023edge, he2020collabovr, ke2023collabvr, leng2019energy, feng2024cicero, zhao2020deja, meng2020coterie, bhojan2020cloudygame}. 
Instead of performing the entire rendering pipeline in the cloud, the rendering workload is partitioned between the server and the client based on their compute resources. 
The server typically performs the more compute- and memory-intensive tasks, while the client processes time-critical and lightweight tasks for responsive interactions.

\hl{Overall, these prior techniques partition the workload at the pixel level, i.e., offloading the majority of pixel rendering to the cloud.
This is because the workload of the traditional rasterization-based pipeline is directly proportional to the number of rendered pixels.
However, their workload partitioning philosophy is incompatible with the 3DGS pipeline, since 3DGS workloads are dominated by the number of processed Gaussians rather than pixel count, as shown in \mbox{\Sect{sec:bg:pipeline}}.
% Moreover, all prior techniques continue to face bandwidth limitations when targeting higher frame rates or resolutions. 
Thus, designing a dedicated collaborative rendering framework for 3DGS remains an open challenge.}

\subsection{Large-Scale 3DGS pipeline}
\label{sec:bg:pipeline}

We first introduce one key concept in large-scale 3DGS rendering, \textit{hierarchical representation}, and we then describe the general pipeline of large-scale 3DGS algorithms.

\para{Hierarchical Representations.} 
\hl{Compared to small-scale 3DGS algorithms\mbox{~\cite{kerbl20233d, fan2023lightgaussian, fang2024mini, mallick2024taming, feng2024flashgs, gui2024balanced, huang2025seele, wang2024adr}}, large-scale 3DGS algorithms\mbox{~\cite{wu2025blockgaussian, kerbl2024hierarchical, liu2024citygaussian, ren2024octree}} introduce hierarchical representations to manage the vast number of Gaussian ellipsoids, the smallest rendering primitives in 3DGS. 
These hierarchical representations enable level-of-detail (LoD) rendering via LoD search, avoiding unnecessary computation when rendering both local views and global views.
% similar to the role of mipmap in the rasterization pipelines~\cite{kilgard2000practical, akenine2019real}. 

During global view rendering, e.g., bird-eye perspectives, LoD allows the pipeline to render all regions at a coarse level, because rendering fine details introduces computational overhead without gaining any quality improvements.
To do so, multiple small Gaussians at a far distance will be merged as a single large Gaussian for rendering.
On the other hand, during local view rendering, e.g., navigating street blocks, LoD inherently leverages the ``divide-and-conquer'' strategy, i.e., rendering local regions at fine details and remote regions more coarsely via its hierarchical representation.
LoD can easily cull irrelevant Gaussians that are outside the current viewpoint or far away from the current location, thus reducing the computational overhead.}

\para{LoD Tree.}
The left part of \Fig{fig:pipeline} gives an example of a hierarchical representation used to store all Gaussians
% in a data structure known as the
, \textit{LoD tree}, where each level corresponds to a specific LoD. 
Each tree node contains a single Gaussian.
The child nodes represent detailed textures of their parent node.
A LoD tree can be implemented by various tree-like structures, such as an octree~\cite{ren2024octree}, an irregular tree~\cite{kerbl2024hierarchical}, or a shallow tree in which each leaf node contains a flattened list of Gaussians~\cite{liu2024citygaussian, wu2025blockgaussian}.

In this paper, we describe the most general form of LoD tree, an irregular tree, where each node is one Gaussian with an arbitrary number of child nodes. 
Gaussians at lower levels of the tree represent finer details.
All other tree-like structures are special cases of this representation.

\para{Pipeline.} 
\Fig{fig:pipeline} shows that a general rendering pipeline for large-scale 3DGS consists of four main stages: \textit{LoD search}, \textit{preprocessing}, \textit{sorting}, and \textit{rasterization}. 

\subparagraph{LoD Search.} 
This stage determines the Gaussian points at an appropriate LoD for subsequent rendering stages. 
Specifically, we first traverse the LoD tree from top to bottom. At each node, we assess if the projected dimension of the Gaussian is smaller than the predefined LoD, $\tau^*$, i.e., the pixel dimension, while the projected dimension of its parent node is larger. 
We then gather all the Gaussians that meet this criterion. 
Conceptually, these selected Gaussians form a ``cut'' that separates the top and bottom of the LoD tree.
% , as the red curve shown in \Fig{fig:pipeline}.

\subparagraph{Preprocessing.}
Once we determine this cut, the selected Gaussians on this cut are projected onto the rendering canvas. 
Gaussians outside the view frustum, e.g., 8, are filtered out.

\subparagraph{Sorting.}
The remaining Gaussians are then sorted by depth, from the nearest to the farthest.

\subparagraph{Rasterization.}
The final stage blends the sorted Gaussians onto the image.
This process is performed tile-by-tile.
Each tile first identifies its Gaussians that intersect with itself and forms its own list, as shown in \Fig{fig:pipeline}.
E.g., tile $T_0$ only intersects Gaussians 1 and 2.
Next, each pixel within a tile performs \textit{$\alpha$-checking}, i.e., calculating the intersected transparency $\alpha_i$ for each Gaussian. 
If $\alpha_i$ falls below a predefined threshold, this pixel would skip that Gaussian for color blending. Otherwise, the Gaussian contributes to the pixel color via weighted blending.
The blending formulation is in~\cite{kerbl20233d}.

\section{Challenges and Opportunities}
\label{sec:ch}

We first describe the challenges in large-scale 3DGS under the contexts of local rendering (\Sect{sec:ch:local}) and remote rendering (\Sect{sec:ch:remote}), separately.
We then explain the insights that can be exploited to address those challenges (\Sect{sec:ch:op}).

\subsection{Challenges in Local Rendering}
\label{sec:ch:local}

\begin{figure}[t]
    \centering
    \includegraphics[width=\columnwidth]{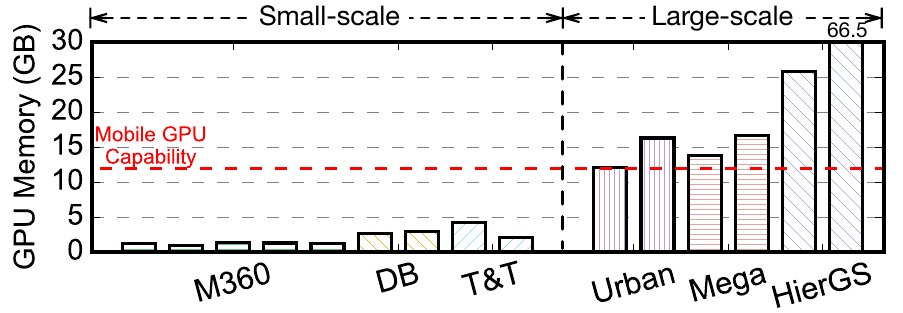}
    \caption{GPU memory footprint trends with scene scale. Runtime numbers are measured across six datasets. Small-scale datasets: T\&T~\cite{Knapitsch2017}, DB~\cite{hedman2018deep}, and M360~\cite{barron2022mipnerf360}. Large-scale datasets: Urban~\cite{lin2022capturing}, Mega~\cite{turki2022mega}, and HierGS~\cite{kerbl2024hierarchical}.}
    \label{fig:memory_pressure}
\end{figure}

\para{Memory Pressure.}
The first challenge in local rendering is the memory pressure imposed by the massive scale of large-scene 3DGS models.
\Fig{fig:memory_pressure} shows the runtime GPU memory footprint across scenes in different datasets~\cite{barron2022mipnerf360, hedman2018deep, Knapitsch2017, kerbl2024hierarchical, lin2022capturing, turki2022mega}.
As the rendering scene scales from small to large, memory usage grows drastically and quickly drains the capacity of mobile GPUs.
All scenes from large-scale datasets exceed the memory capacity of mainstream VR devices~\cite{questprospec, quest3spec, htcvivespec, visionprospec}, which are often less than 12 GB.
Specifically, a scene from HierGS~\cite{li2023matrixcity} even exceeds 66~GB.
However, prior architectural designs~\cite{feng2025lumina, ye2025gaussian, feng2024potamoi, lee2024gscore, li2025uni, lee2025vr, lin2025metasapiens, durvasula2025arc, he2025gsarch} have primarily focused on small scenes, largely ignoring the scalability challenges posed by large-scale 3DGS contents.
Without addressing this bottleneck, it is infeasible to achieve infinite-scale 3DGS rendering in the foreseeable future.

\begin{figure}[t]
\centering
\begin{minipage}[t]{0.48\columnwidth}
  \centering
  \includegraphics[width=\columnwidth]{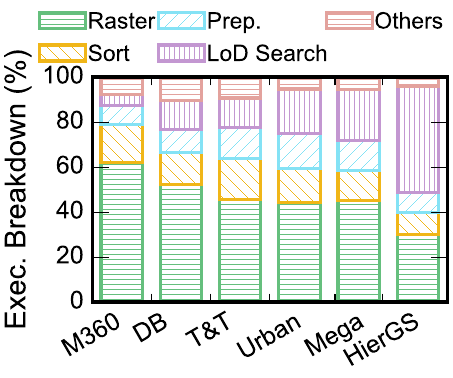}
  \caption{The end-to-end execution breakdown of \textit{local rendering} on a mobile Ampere GPU~\cite{orinsoc}. ``Others'': time on sensor tracking and display.}
  \label{fig:exec_time}
\end{minipage}
\hspace{2pt}
\begin{minipage}[t]{0.48\columnwidth}
  \centering
  \includegraphics[width=\columnwidth]{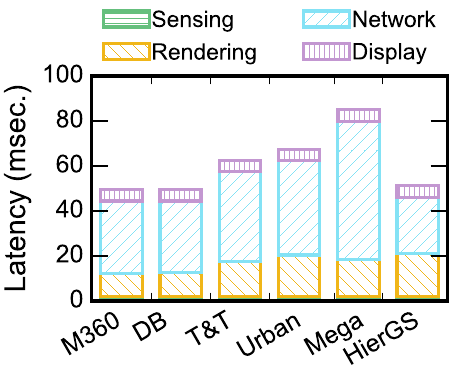}
  \caption{The end-to-end execution breakdown of \textit{remote rendering}. Data transmission is the major bottleneck under 90~FPS VR resolution.}
  \label{fig:remote_time}
\end{minipage}
\end{figure}

\para{Bottleneck Shift.}
Another key observation in large-scale 3DGS rendering is that, as the scene complexity increases, the computational bottleneck shifts from rasterization to LoD search. 
Quite a few studies~\cite{lee2024gscore, feng2025lumina, ye2025gaussian, lin2025metasapiens} propose dedicated accelerators for rasterization, which dominates the execution time in small-scale scenes in \Fig{fig:exec_time}. 
However, with increasing scene complexity, the cost of LoD search, i.e., identifying which Gaussians should be rendered, increases rapidly and begins to dominate the overall execution.

As shown in \Fig{fig:exec_time}, the relative execution time of LoD search increases with the scene size on a Nvidia mobile Ampere GPU~\cite{orinsoc}, accounting for up to 47\% of the end-to-end latency in large-scale 3DGS scenes.
In contrast, the relative time of rasterization does not grow with scene scales, because, with LoD search, the number of Gaussians that can contribute to the final frame plateaus. 
Therefore, without effectively supporting LoD search, it remains infeasible to achieve real-time rendering on mobile devices.

\subsection{Challenges in Remote Rendering}
\label{sec:ch:remote}

The main challenge in remote rendering is the network transmission. In the following, we explain the transmission challenges in both \textit{video streaming} and \textit{collaborative rendering}.

\para{Video Streaming.}
Prior studies~\cite{boos2016flashback, lai2017furion, kanter2018graphics} have shown that human visual system is very sensitive to both latency and resolution.
To deliver a smooth user experience, VR applications often require high-frame-rate ($>$ 90~FPS) and high-resolution stereo video streams (e.g., a middle-tier headset, Meta Quest 3, features $2064\times2208$ pixels per eye~\cite{quest3spec}).

Under this resolution requirement, \Fig{fig:remote_time} shows the breakdown of the end-to-end latency across 3DGS datasets.
The experiential setup is detailed in \Sect{sec:exp}.
Our results show that the overall end-to-end latency is dominated by the data transmission overhead due to video streaming.

\begin{figure}[t]
\centering
\begin{minipage}[t]{0.48\columnwidth}
  \centering
  \includegraphics[width=\columnwidth]{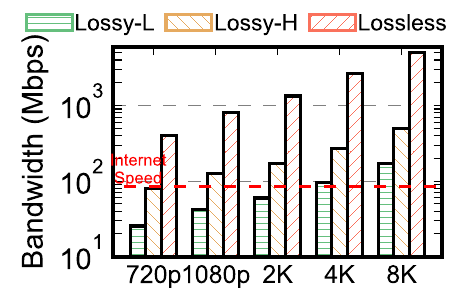}
  \caption{The trend of network bandwidth demand with the increasing resolution. Red dashed line shows the average US household internet speed in 2025~\cite{networkspec}.}
  \label{fig:network_req}
\end{minipage}
\hspace{2pt}
\begin{minipage}[t]{0.48\columnwidth}
  \centering
  \includegraphics[width=\columnwidth]{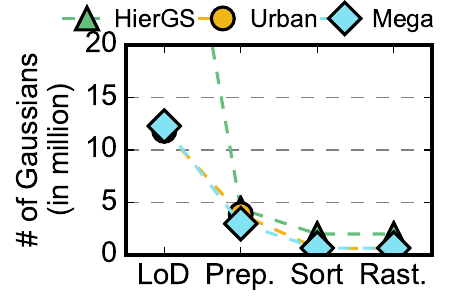}
  \caption{The runtime memory demand varies across different stages. We use the number of involved Gaussians as a proxy for memory demand.}
  \label{fig:memory_req}
\end{minipage}
\end{figure}

In addition, \Fig{fig:network_req} demonstrates that the network pressure would further escalate as the resolution demand increases.
Here, we show different compression schemes, lossy and lossless video compression using H.256~\cite{h265}, with ``L'' and ``H'' denoting low- and high-quality settings.
we show that streaming high-quality videos under VR settings well surpass today's network bandwidth~\cite{networkspec}.
Even with recent efforts on multi-view video compression~\cite{djelouah2019neural, tang2024offline, yang2018cross}, conventional video streaming is still not a scalable solution.
% for high-resolution, high-frame-rate VR rendering.

\para{Collaborative Rendering.}
Recent studies~\cite{meng2020coterie, xie2021q, leng2019energy, xu2023edge} have explored various techniques that leverage both the computational power of remote servers and local clients for conventional VR rendering.
% The core of these works is to partition the overall rendering workload (e.g., rendering assets) between the server and the client based on their computational power.
However, unlike conventional mesh-based rendering, 3DGS scenes are built by a set of Gaussian points. 
Currently, there is no mechanism to manage these Gaussians and determine which subset should be delegated to local rendering. 
Meanwhile, in those studies, the primary workloads of the scene are still rendered in the cloud; thus, the majority of pixels must be transmitted to the client for display. 
As a result, data transmission remains the dominant bottleneck in such collaborative rendering setups.

To fill this gap, we propose the \textit{first} Gaussian-based asset streaming technique that dynamically transmits Gaussians from the cloud to the client. 
Our approach is inherently insensitive to resolution and frame rate demands, enabling scalable 3DGS rendering. 
The following \Sect{sec:ch:op} highlights the key insights that we leverage.

\subsection{Key Insights}
\label{sec:ch:op}

\para{Memory Demand.}
\Sect{sec:ch:local} shows that large-scale 3DGS scenes exceed the memory capacity of VR devices, however, we show that the runtime memory demand varies drastically across stages of the 3DGS pipeline.
\Fig{fig:memory_req} measures the memory demand across different stages, using the number of involved Gaussians as a proxy for memory demand.

The initial stage incurs the highest memory footprint, as LoD search potentially needs to filter over all Gaussians in the scene to determine the appropriate LoD.
After the LoD search, the number of Gaussians drops quickly to a point where a mobile GPU can easily accommodate.
This result shows that the entire 3DGS pipeline can potentially be split into two halves, allowing us to offload the high-memory-demand part, i.e., LoD search, to the cloud.

\begin{figure}[t]
\centering
\begin{minipage}[t]{0.48\columnwidth}
  \centering
  \includegraphics[width=\columnwidth]{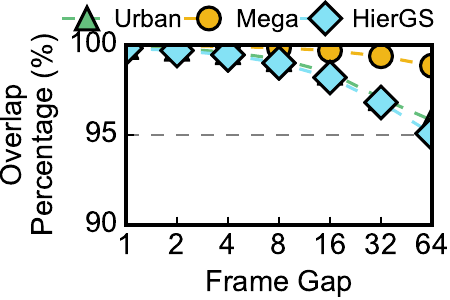}
  \caption{The temporal similarity between adjacent frames under a 90~FPS VR scenario.}
  \label{fig:profile_overlap}
\end{minipage}
\hspace{2pt}
\begin{minipage}[t]{0.48\columnwidth}
  \centering
  \includegraphics[width=\columnwidth]{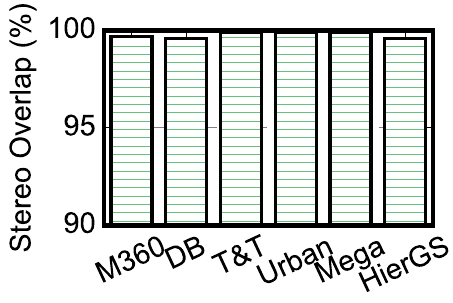}
  \caption{The stereo similarity between the left-eye and right-eye images in VR.}
  \label{fig:stereo_overlap}
\end{minipage}
\end{figure}

\begin{figure*}[t]
    \centering
    \includegraphics[width=\textwidth]{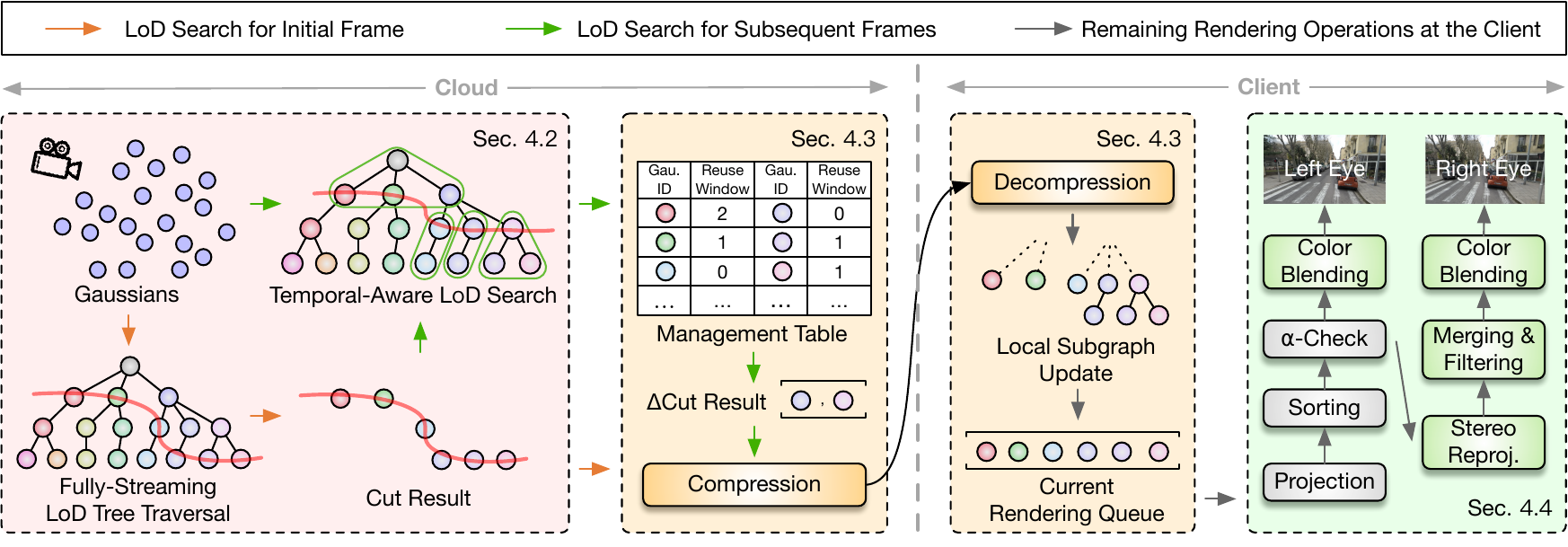}
    \caption{Overview of \proj workflow. \proj offloads the memory-intensive LoD search to the cloud while the client side executes the remaining operations. 
    On the cloud side, we propose a temporal-aware LoD search that exploits the temporal similarity across frames to accelerate LoD tree traversal (\Sect{sec:algo:search}). 
    On the client side, we propose a stereo rasterization algorithm that leverages the geometric relationship between Gaussians and the stereo camera to avoid redundant computation (\Sect{sec:algo:stereo}).
    Lastly, a Gaussian management system coordinates the cloud–client interface and alleviates bandwidth bottlenecks (\Sect{sec:algo:mem}).}
    \label{fig:overview}
\end{figure*}

\para{Temporal Similarity.}
Meanwhile, we also find that the set of Gaussians selected after LoD search, the ``cut'' in \Fig{fig:pipeline}, exhibits a strong temporal similarity across adjacent frames.
Our experiment in \Fig{fig:profile_overlap} shows the overlap ratio of Gaussians after LoD search using HierGS dataset~\cite{kerbl2024hierarchical}. 
Here, we simulate a 90 FPS VR real-time rendering scenario.
Experiments show that 99\% of the Gaussians after LoD search remain unchanged between two consecutive frames. 
Even with a frame gap exceeding 64, there is still $>$95\% of the Gaussians that are identical.
High temporal similarity indicates that substantial computations in LoD search are redundant.

\para{Stereo Similarity.}
In addition to temporal similarity across frames, the two frames rendered for the left and right eyes in VR also exhibit strong similarity. 
\Fig{fig:stereo_overlap} shows the percentage of overlapping pixels between the two frames across different datasets. 
To quantify this, we warp the left-eye image to the right-eye view using a technique similar to Cicero~\cite{feng2024cicero}.
Our results show that fewer than 1\% of the pixels are non-overlapping between the left and right eyes. 

However, there is one caveat: 3DGS rendering is \textit{view-dependent}, meaning that the same physical point may appear with different colors at different viewing directions (e.g., specular reflection). 
Thus, directly warping pixels from the left eye to the right eye could introduce noticeable artifacts.

% \para{Summary.}
% Here, we summarize the key insights:
% 1) LoD search dominates the memory usage along the pipeline, while the demands of later stages drop sharply.
% 2) Rendering across consecutive frames and stereo views shows strong temporal and stereo similarity.

% \begin{itemize}
%     \item LoD search dominates the memory usage along the pipeline, while the demands of later stages drop sharply.
%     \item Rendering across consecutive frames and stereo views shows strong temporal and stereo similarity.
% \end{itemize}

\section{Framework}
\label{sec:algo}

To address the challenges in both local and remote rendering in \Sect{sec:ch}, we introduce our collaborative rendering framework, \proj. Our key idea is to orchestrate the compute resources of both the cloud and the client while minimizing the data communication between these two.

\hl{We first give an overview of three key contributions in \mbox{\proj} (\mbox{\Sect{sec:algo:ov}}).
% then we describe the individual components in our framework.
We then explain our temporal-aware LoD search on the cloud side in \mbox{\Sect{sec:algo:search}}.
Next, we show how we transmit our data from the cloud to the client in \mbox{\Sect{sec:algo:mem}}.
Lastly, we introduce our novel stereo rendering pipeline on the client side in \mbox{\Sect{sec:algo:stereo}}.}

\subsection{Overview}
\label{sec:algo:ov}

\para{Design Principle.}
To ensure broad applicability and seamless integration into existing systems, our design follows two key principles:
1) our algorithm on the cloud side should run out-of-the-box on existing server GPUs without any hardware modifications; 2) our algorithm on the client side should be easily integrated into existing 3DGS accelerators with only minimal hardware augmentations.

\para{Idea.}
Guided by our design principles, \proj exploits the three key insights in \Sect{sec:ch:op}.
First, by leveraging the memory demand characteristics, we offload the most memory-intensive LoD search to the cloud, while leaving the remaining stages to be executed by the dedicated accelerators on the client side.
Second, we leverage the temporal similarity of the intermediate results after LoD search to reduce the per-frame data communication between the cloud and client.
Lastly, our rendering pipeline exploits the stereo similarity of the binocular view to avoid redundant computation.
Note that, \textit{we offload only LoD search, not preprocessing, to the cloud, so that the client can render any viewport near the current position with no extra data traffic}.
This way, we can accommodate the rapid head rotations in VR/AR~\cite{blandino2021head, hendicott2002head, wang2023effect}.

\para{Workflow.}
\Fig{fig:overview} shows the overall workflow of our pipeline.
First, the cloud side executes the LoD search.
Here, we propose a GPU-efficient \textit{temporal-aware LoD search} that leverages the temporal similarity across frames to avoid unnecessary tree node accesses.
Specifically, \proj processes the initial frame and the subsequent frames separately.

\begin{figure*}[t]
    \centering
    \includegraphics[width=\textwidth]{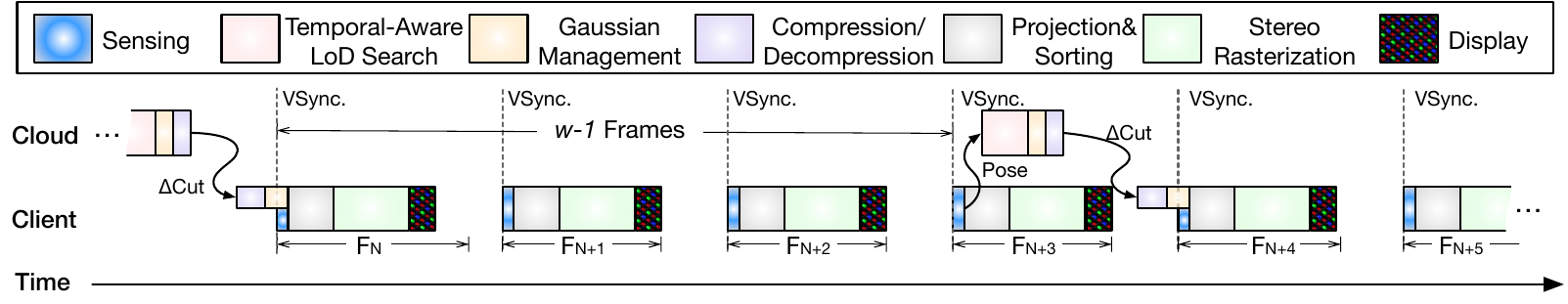}
    \caption{The timing diagram of our execution flow. The latency on the cloud side can be hidden by the latency of multiple locally rendered frames. Only client-side operations are on the critical path.}
    \label{fig:exec_flow}
\end{figure*}

For the initial frame, \proj performs a full LoD search, i.e., a LoD tree traversal, in the cloud to find the Gaussians at the appropriate LoD, the ``cut'' (\Fig{fig:pipeline}), at the current pose.
To avoid irregular memory access that commonly exists in tree traversal~\cite{pinkham2020quicknn, feng2022crescent, xu2019tigris, feng2020mesorasi, feng2025streamgrid}, we propose a \textit{fully streaming LoD tree traversal} that is highly parallelizable on GPUs.

For subsequent frames, \proj leverages the temporal similarity of the ``cut'' results across adjacent frames.
Instead of performing a complete tree traversal from the root node, our temporal-aware LoD search in \Sect{sec:algo:search} performs lightweight local updates by searching only within relevant subtrees to update the cut result of the current frame.
% Further explanations of our algorithm are given in \Sect{sec:algo:search}.

Once we obtain the cut result from the LoD search, we transmit the necessary data to the client side.
% While sending the complete results for every frame is possible, doing so would incur substantial communication overhead.
Our \textit{runtime Gaussian management} system in \Sect{sec:algo:mem} leverages the temporal similarity across adjacent frames.
Instead of sending a complete cut result, it compresses and transmits only the non-overlapped Gaussians in the cut result.
Thus, we reduce the data transfer between the cloud and the client.

Once the client receives Gaussians from the cloud, the client decompresses the data and updates its local subgraph that only reserves the recently used Gaussians for the client.
The subgraph update is co-designed with the Gaussian management system in the cloud so that both ends have a consistent view of which Gaussians the client has (\Sect{sec:algo:mem}).

Lastly, our \textit{stereo rasterization} pipeline uses the Gaussians from the local subgraph to render left- and right-eye images.
Instead of rendering two separate frames for these two eyes, our pipeline exploits the geometric relationship between Gaussians and the stereo camera, allowing these two frames to share most of the computations while still producing bit-accurate images as if rendering them separately.
We further explain our rasterization algorithm in \Sect{sec:algo:stereo}.

\para{Timing.}
\Fig{fig:exec_flow} summarizes the timing of our execution flow. 
On the cloud side, we first perform the temporal-aware LoD search and update the Gaussian management system. 
The cloud then compresses the Gaussians that need to be transmitted and sends them to the client over the network.
On the client side, we first decompress the data upon arrival. 
Meanwhile, the client continuously traverses the subgraph and the remaining stages of the 3DGS rendering.
Once rendering completes, the new image is displayed at the next VSync arrival.
Following the convention of large-scale 3DGS~\cite{kerbl2024hierarchical}, LoD search is executed only once every $w$ frames, e.g., $w = 4$.
In \Fig{fig:exec_flow}, the latency on the cloud side can be hidden by multiple locally rendered frames.
Importantly, the client can also continue rendering subsequent frames without waiting for cloud data with a negligible quality sacrifice.
Thus, only client-side operations are on the critical path.
% The rendering quality comparison is shown in \Sect{sec:eval:acc}.

\subsection{Temporal-Aware LoD Search}
\label{sec:algo:search}

% In this subsection, 
We first explain our fully-streaming tree traversal to accelerate the initial frame.
Next, we describe how temporal-aware LoD search can leverage previous results to avoid unnecessary tree node traversal in subsequent frames.
This paper only focuses on the tree traversal acceleration.
For the details on the LoD tree construction, please refer to HierGS~\cite{kerbl2024hierarchical}.

\begin{figure}[t]
    \centering
    \includegraphics[width=\columnwidth]{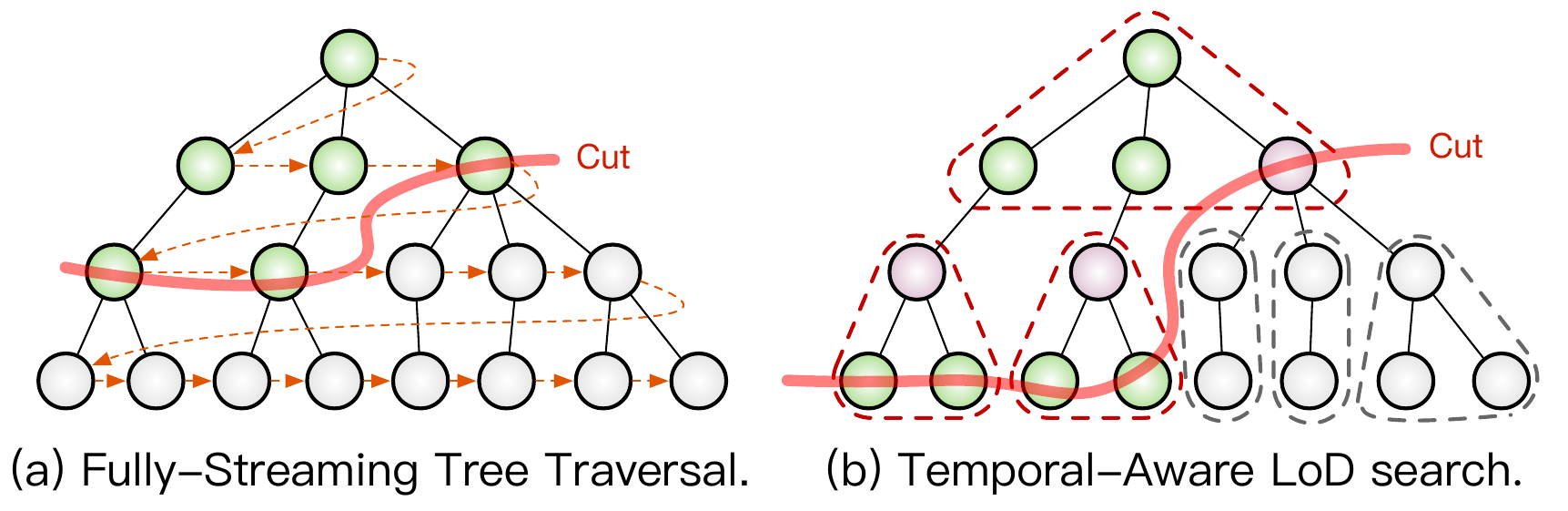}
    \caption{The illustration of fully-streaming LoD tree traversal and temporal-aware LoD search.}
    \label{fig:tree_traversal}
\end{figure}

\para{Fully-Streaming LoD Tree Traversal.} \Fig{fig:tree_traversal}a shows our \textit{fully-streaming LoD tree traversal} for initial frame. 
A LoD tree is an irregular tree, where each node represents one Gaussian with an arbitrary number of child nodes. 
A normal tree traversal would lead to irregular DRAM accesses~\cite{pinkham2020quicknn, feng2022crescent}.
The goal of our algorithm is to achieve high parallelism on GPUs while minimizing unnecessary tree node visits.
Rather than relying on the inherent parent-child relations in the original LoD tree (denoted by solid arrows), we augment the LoD tree with connections that enable tree traversal in breadth-first order (denoted by orange dashed arrows).

During the GPU execution, different GPU warps are assigned the same amount of workload, i.e., a block of tree nodes. 
Each block is then evenly distributed among threads within a warp to ensure a balanced workload across GPU threads. 
We design each block of nodes to be small enough to fully reside in GPU shared memory, so that our algorithm can streamingly process those tree nodes and avoid irregular DRAM accesses.
The workload assignment is dynamically dispatched whenever a GPU warp becomes available at runtime. 
The tree traversal terminates once a clean cut separates the top and bottom of the LoD tree (see red curve in \Fig{fig:tree_traversal}a).
This way, our algorithm processes only the green nodes while completely skipping the grey ones, thus avoiding redundant computation and irregular DRAM access.

\para{Temporal-Aware LoD Search.} For tree traversal of subsequent frames, we introduce a \textit{temporal-aware LoD search}.
Our algorithm first partitions the entire LoD tree into multiple subtrees offline, while preserving the hierarchical relationships within the LoD tree.
\Fig{fig:tree_traversal}b highlights the individual subtrees in dashed blocks.
Here, we only show a two-level subtree partitioning for illustration purposes; our actual implementation applies multi-level partitioning to the LoD tree.

Given the cut result from the previous frame (highlighted in pink), our algorithm first identifies the subtrees to which each Gaussian in the cut belongs.
\Fig{fig:tree_traversal}b highlights these subtrees in red dashed blocks.
Then, our algorithm only traverses those identified subtrees instead of all subtrees initially.
In GPU implementation, each subtree is assigned to a separate GPU warp for local subtree traversal.
The subtree partitioning is performed offline and guarantees that each subtree is approximately equal in size, ensuring balanced workload distribution across GPU warps.
If searching the local subtree cannot obtain the complete ``cut'' result, i.e., no clean cut for this subtree, we then search its corresponding top-tree or subtrees to complete the cut finding. 
Note that, the results from our temporal-aware LoD search are \textit{bit-accurate} compared to those of the original full-tree traversal.
In \Fig{fig:lod_speedup}, we compare our tree traversal performance against prior designs and achieve up to 52.7$\times$ speedup.

\subsection{Runtime Gaussian Management}
\label{sec:algo:mem}

We next describe how \proj transfers only the necessary Gaussians to the client at runtime to minimize the data transfer between the cloud and client.
Our system design should guarantee two key properties: 1) the cloud and client share a consistent view of which Gaussians are currently stored on the client; and 2) obsolete Gaussians should be removed from the client on the fly to alleviate memory pressure.
We now describe the system design on both sides.

\para{Cloud Side.}
Our system in the cloud maintains a management table that tracks the set of Gaussians currently stored on the client (\Fig{fig:overview}).
For each Gaussian, our management system maintains a reuse window, $w_r$, which is used to represent the number of frames since this Gaussian was last included in a cut result.
When a new cut result is generated via LoD search, the cloud iterates over Gaussians in the cut result and updates those that do not exist in the management table.
Those newly encountered Gaussians are then gathered into a group (called a $\Delta$cut) and transmitted to the client.

Meanwhile, our system can also remove obsolete Gaussians for the client.
Here, the cloud and client share the same reuse threshold, denoted $w_r^*$. 
Here, we set $w_r^*$ to be 32.
After the table is updated, both the cloud and client iterate through their tables and remove any Gaussians whose reuse window $w_r$ exceeds the threshold $w_r^*$. 
Those Gaussians are considered obsolete and removed from the management table.
The overall idea is similar to garbage collection~\cite{lieberman1983real, appel1989simple}.

\para{Client Side.}
Similar to the cloud side, each time the client receives a $\Delta$cut, the client updates its local LoD subgraph with new Gaussians. 
The client side is relatively straightforward.
After completing the insertion, the client performs the same removal process as the cloud to discard obsolete Gaussians whose reuse window exceeds $w_r^*$. 
Finally, to render the next frame, the client iterates over the list of Gaussians on the client side and generates a rendering queue for Gaussians with an appropriate LoD.

% Similar to the cloud side, each time the client receives a $\Delta$cut, the client updates its local LoD subgraph with new Gaussians. 
% To avoid maintaining the entire LoD tree with high memory overhead, the client uses a lightweight hash table to store and manage Gaussians.
% To insert a Gaussian into the hash table, the client needs to update the entry for the Gaussian itself, as well as its associated parent and child nodes. 
% Each Gaussian entry includes attributes such as its parent ID, child IDs, spherical harmonic (SH) coefficients, and other attributes for rendering.
% After completing the insertion, the client performs the same removal process as the cloud to discard obsolete Gaussians whose reuse window exceeds $w_r^*$. 
% Finally, to render the next frame, the client iterates over the updated hash table and generates a rendering queue for Gaussians with an appropriate LoD.

\para{Compression.} 
While our management system already reduces the number of Gaussians that need to be transmitted, we further apply a compression technique to lower the data transmission rate.
Following the approach of prior works~\cite{papantonakis2024reducing, lee2024compact}, we compress different Gaussian attributes independently.
For the most storage-intensive SH coefficients, we adopt vector quantization, similar to Compact3DGS~\cite{lee2024compact}.
Other attributes, such as position and scale, which together account for only a small fraction of the overall storage, are encoded using a 16-bit fixed-point representation with negligible quality loss (\Sect{sec:eval:acc}).
Here, we claim no contribution.

\subsection{Stereo Rasterization}
\label{sec:algo:stereo}

\Sect{sec:ch:op} shows the strong similarity between left- and right-eye images.
This subsection describes our stereo rendering pipeline, which leverages the stereo similarity between the two eyes to reduce the computations on the client side.

\para{Motivation.}
While prior studies~\cite{feng2024cicero, chaurasia2020passthrough+, vona2025comparing} have proposed methods to exploit stereo similarity between two eye images, these techniques have two key limitations.
First, existing methods rely on a high-fidelity depth map to perform accurate warping; however, the depth maps produced by 3DGS are often unreliable.
Second, directly warping pixels from the left eye to the right eye compromises the view-dependent characteristic of 3DGS.
It often produces less photo-realistic images, as mentioned in prior work~\cite{feng2024cicero}.

To address these issues, we propose a triangulation-based technique that skips redundant computation in the remaining stages, i.e., preprocessing, sorting, and rasterization, while preserving \textit{bit-accurate} results in 3DGS.

\begin{figure}[t]
    \centering
    \includegraphics[width=\columnwidth]{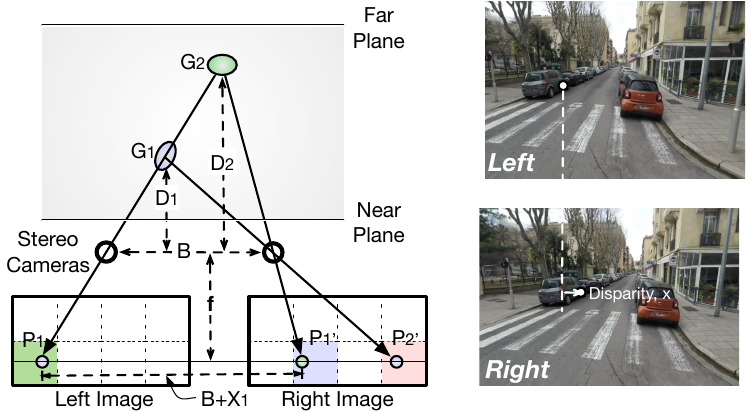}
    \caption{The intuition behind our stereo rasterization, which leverages the triangulation technique~\cite{hartley2003multiple, szeliski2010computer}. 
    For each Gaussian, once we determine the pixels it intersects in the left-eye image, we can directly compute, via triangulation, the corresponding pixel locations it will contribute to in the right-eye image, without preprocessing and sorting. }
    \label{fig:stereo_vision}
\end{figure}

\begin{figure*}[t]
    \centering
    \includegraphics[width=\textwidth]{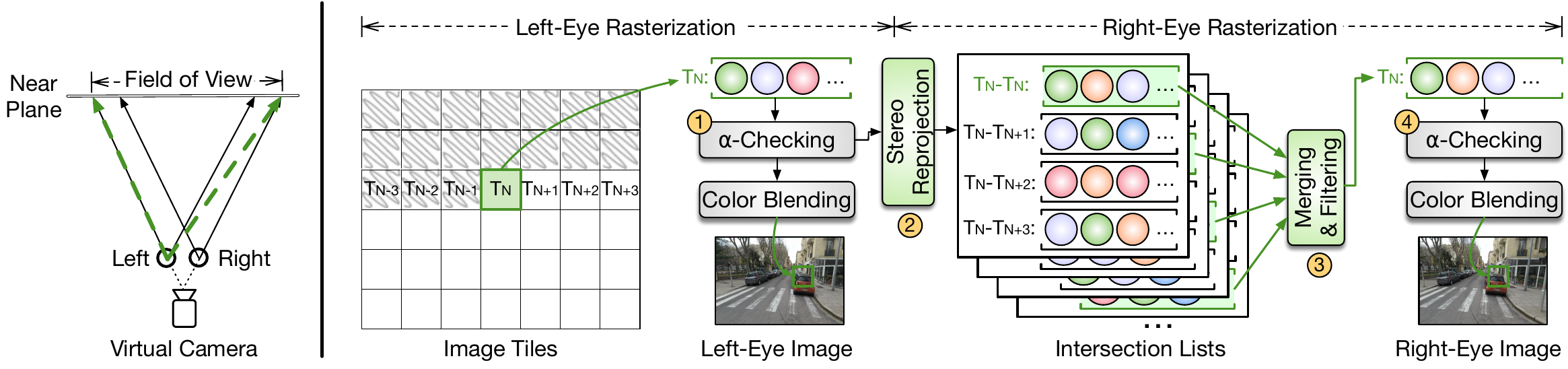}
    \caption{Overview of stereo rasterization. Left: for preprocessing and sorting, we use a wider FoV to cover the FoVs of both eyes. Right: for rasterization, we first perform a standard rasterization to render the left-eye image; for the right-eye image, instead of reprocessing all Gaussians, we leverage the geometric relationship between Gaussians and the stereo camera, and map those contributing Gaussians to the right view via triangulation. Thus, we largely reduce the redundant computations.
}
    \label{fig:stereo_rasterization}
\end{figure*}

\para{Intuition.}
We first give a toy example in \Fig{fig:stereo_vision} to show the intuition behind our algorithm: \textit{given the Gaussian relative depth to the left camera and the geometry of the stereo camera, we can directly compute which pixel in the right image would be contributed by this Gaussian point.}

\Fig{fig:stereo_vision} shows a stereo camera with a baseline of $B$, defined as the horizontal distance between the left and right cameras in a VR headset.
Both cameras have the same focal length, $f$.
Consider a pixel $P_1$ in the left-eye image, which is contributed to by two Gaussians, $G_1$ and $G_2$.
The depths of $G_1$ and $G_2$ to the camera center are $D_1$ and $D_2$, respectively.

Based on \textit{triangulation}~\cite{hartley2003multiple, szeliski2010computer}, a widely-known process in computer vision, we calculate the \textit{disparity}, $X_1 = P_1' - P_1$, i.e., the horizontal displacement between the two corresponding pixels $P_1'$ and $P_1$ in the left and right images as, $X_1 = Bf/D_1.$
% \begin{equation}
% \label{eq:triangulation}
% X_1 = \frac{B f}{D_1}.
% \end{equation}
Similarly, the disparity for $G_2$ is given by $X_2 = Bf/D_2$.
With these disparity results, we can determine, in the right image, which pixels each Gaussian will contribute to, without re-running the preprocessing and sorting stages.
% Note that the corresponding pixel for $G_1$ has a larger disparity than that of $G_2$, since disparity is inversely proportional to depth.
In 3DGS rendering, a near plane and a far plane are defined to avoid rendering artifacts~\cite{kerbl20233d}.
Given the near-plane distance, the maximum disparity in a typical VR setup is bounded within 16 pixels, since disparity is inversely proportional to depth.

\para{Algorithm.}
In \Sect{sec:bg:pipeline}, we explain that 3DGS pipelines render in a tile-by-tile fashion, specifically, $4\times4$ granularity.
Our algorithm also adapts the tile-based rendering and maps every Gaussian contributing to a tile in the left-eye image to its corresponding tile in the right-eye image.
For example, $G_1$, which contributes to the green tile in the left image, is mapped to the pink tile in the right-eye image in \Fig{fig:stereo_vision}. 
We now describe the changes in the rendering pipeline.

\subparagraph{Preprocessing\&Sorting.}
The left part of \Fig{fig:stereo_rasterization} shows our key modification to the preprocessing and sorting stages.
That is to allow the left- and right-eye images to share these computations of these two stages, given their highly similar fields of view (FoVs).
To do that, we place a virtual camera slightly behind both eyes to determine the common FoV between them.
Then, we derive the left-eye FoV that fully covers the common region (the green bashed FoV), and perform preprocessing and sorting on this FoV to avoid repetitively computing the preprocessing and sorting twice.

\subparagraph{Stereo Rasterization.}
The right part of \Fig{fig:stereo_rasterization} shows our key contribution, \textit{bit-accurate} stereo rasterization pipeline:

\circnum{1} 
We first render the left‑eye image following the standard rasterization process.
For a given tile $T_N$, each pixel iterates over all Gaussians in the $T_N$ list, which is a sorted list after sorting (see \Fig{fig:pipeline}). 
For each Gaussian, each pixel first performs an $\alpha$-check: if the transmittance exceeds the threshold $\alpha^{*}$, the Gaussian’s color is blended into the pixel; otherwise, the Gaussian is skipped.
The final pixel value is obtained after processing all Gaussians in the list.

\circnum{2}
If a Gaussian passes an $\alpha$-check, it definitely contributes to the right-eye image.
We then apply triangulation, as described in \Fig{fig:stereo_vision}, to transform this Gaussian to the right-eye view.
Based on the computed disparity, we can determine which tile in the right-eye image this Gaussian intersects, and insert this Gaussian into the corresponding list.
Based on the maximum disparity (16 pixels), this Gaussian will be inserted into one of four lists, i.e., $T_N$-$T_N$, $T_N$-$T_{N+1}$, $T_N$-$T_{N+2}$, and $T_N$-$T_{N+3}$.
$T_N$-$T_{N+1}$ stands for the Gaussians from tile $T_N$ in the left image that would contribute to tile $T_{N+1}$ in the right image.
Each tile maintains its four intersection lists.

\circnum{3}
To render tile $T_N$ in the right-eye image, we first need to identify which Gaussians intersect with this tile.
Based on the triangulation, the intersected Gaussians can only come from four lists: $T_{N-3}$-$T_N$, $T_{N-2}$-$T_N$, $T_{N-1}$-$T_N$, and $T_N$-$T_N$.
The complete intersection set for $T_N$ is obtained by merging these four lists.
Since each list is already sorted, we can skip re-sorting these four lists.
Instead, we directly merge them and remove duplicate ones.
This process is analogous to the merge phase of merge sort, but with four pre-sorted lists.

\circnum{4}
Once we obtain the complete and sorted intersection list for $T_N$, we can render tile $T_N$ of the right-eye image following the same procedure as for the left-eye image.
Since our pipeline integrates the same list of Gaussians as the original pipeline, the rendering result is bit-accurate.

In this process, the Gaussians processed by the right-eye image have already passed the $\alpha$-check; thus, our stereo rasterization inherently avoids part of rasterization workload for right eyes and achieves 1.4$\times$ speedup on mobile GPUs.
% \Sect{sec:eval:perf} shows that our software implementation achieves 1.4$\times$ speedup on mobile GPUs.

\section{Architectural Support}
\label{sec:arch}

\begin{figure}[t]
    \centering
    \includegraphics[width=\columnwidth]{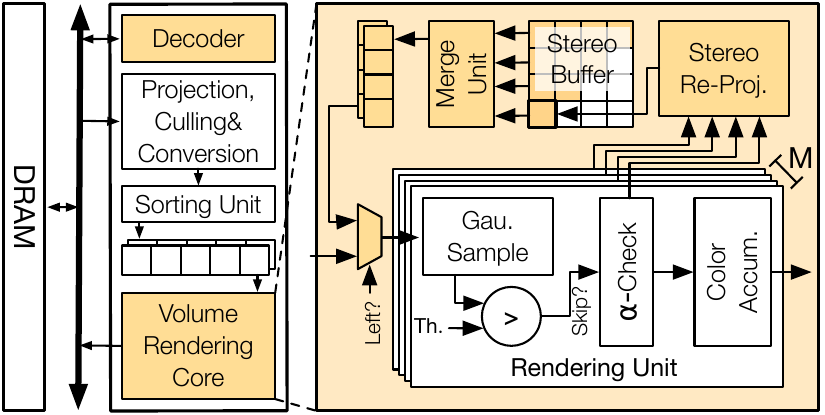}
    \caption{The overview architecture design. We augmented the basic architecture, GSCore~\cite{lee2024gscore}, to support decompression and stereo rasterization (colored in yellow).}
    \label{fig:arch}
\end{figure}

Supporting stereo rasterization in \Sect{sec:algo:stereo} on existing 3DGS accelerators requires only minimal hardware augmentation.
In this section, we use GSCore~\cite{lee2024gscore} as an example to illustrate the necessary modifications.
Other architectures can adapt our technique in a similar fashion.

\para{Overview.}
\Fig{fig:arch} illustrates the overall pipelined architecture of \proj, built upon GSCore~\cite{lee2024gscore}, where the projection unit, sorting unit, and volume rendering core (VRC) are pipelined to render different tiles within a frame.
We introduce two hardware augmentations.
First, the VRC in GSCore is augmented to support stereo rasterization.
Second, we design a lightweight decoder for compressed Gaussian attributes.
The decoder, equipped with a codebook buffer to store codewords, is used to map encoded indices back to the original Gaussian attributes~\cite{vasuki2006review, gray1984vector}.
As the decoder design is straightforward, this section focuses primarily on the VRC.

\para{Support for Stereo Rasterization.}
\Fig{fig:arch} highlights our modified components in colors.
The basic VRC consists of $M$ rendering units (RUs).
Each RU is responsible for rendering one pixel.
For each Gaussian, its attributes are first broadcast to all RUs.
Based on the result of the $\alpha$-check, each RU determines whether the Gaussian contributes to its pixel and, if so, blends it into the accumulated color.

Meanwhile, all the results of the $\alpha$-check are forwarded to our augmented stereo re-projection unit (SRU).
If any pixel within the tile integrates this Gaussian, then the SRU would re-project this Gaussian into the right-eye view.
Based on the re-projected disparity, the SRU would write the Gaussian into the corresponding list in the stereo buffer.

\begin{figure}[t]
    \centering
    \includegraphics[width=\columnwidth]{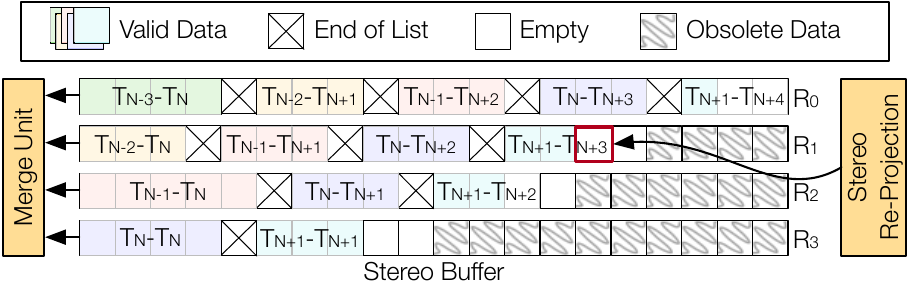}
    \caption{The data layout of the stereo buffer. Different colors denote different tiles in the left-eye image. Our design adopts a line buffer structure to eliminate the bank conflict.}
    \label{fig:buffer}
\end{figure}

Our stereo buffer adopts the classic line buffer design from image processing~\cite{whatmough2019fixynn, chi2018soda, hegarty2014darkroom, ujjainkar2023imagen}, as shown in \Fig{fig:buffer}.
To avoid the bank conflicts, each row stores a single disparity category.
For instance, row $R_0$ stores Gaussians whose disparity is greater than 3 tiles (i.e., 12 pixels) between the left and right eyes.
Here, $T_{N-3}$-$T_{N}$ denotes the Gaussians from tile $T_{N-3}$ in the left image that would contribute to tile $T_{N}$ in the right image.
Each time, SRU writes a Gaussian to one of the rows based on the disparity result.
Meanwhile, the merge unit sorts the current four lists for $T_N$ in the right image, i.e., $T_{N-3}$-$T_{N}$, $T_{N-2}$-$T_{N}$, $T_{N-1}$-$T_{N}$, and $T_{N}$-$T_{N}$, by reading the head entries of the four rows and selecting the minimum.
Each row is designed as a circular buffer to maximize utilization.

\para{Pipelining.}
Similar to GSCore~\cite{lee2024gscore}, our architectural design pipelines the three stages, preprocessing, sorting, and rasterization, and renders image tiles in row-major order.
In our stereo rasterization, we sequentially render the corresponding tiles in the left-eye and right-eye images.
Note that, right-eye tiles begin to render after left-eye tile rendering, starting from the fourth tile.
The first three tiles in the right-eye images are rendered independently.

\section{Experimental Setup}
\label{sec:exp}

\para{Hardware Implementation.}
We develop a RTL implementation of \proj clocked at 1 GHz, where the basic configuration is similar to GSCore~\cite{lee2024gscore}.
\proj consists of four projection units, four hierarchical sorting units, and eight VRCs.
Each VRC consists of $4\times4$ rendering units and a 16~KB feature buffer.
We augment each VRC with one stereo reprojection unit, one merge unit, and a 16~KB stereo buffer banked at 4~KB granularity.
In addition, a 144~KB global double buffer is used to store the intermediate data of the pipeline.
Our RTL design is implemented via Synposys synthesis and Cadence layout tools in TSMC 16nm FinFET technology. 
SRAMs generated by an Arm memory compiler. 
Power is simulated using Synopsys PrimeTimePX, with fully annotated switching activity.
The DRAM is modeled after 4 channels of Micron 16~Gb LPDDR3-1600 memory~\cite{micronlpddr3}.
DRAM energy is calculated using Micron's System Power
Calculators based on the memory traffic~\cite{microdrampower}.
The numbers of our RTL designs are then scaled down to 8 nm node using DeepScaleTool~\cite{stillmaker2017scaling, sarangi2021deepscaletool} to match the mobile Ampere GPU on Nvidia Orin~\cite{orinsoc}.

\para{Area.}
\proj introduces a minimal area overhead compared to the baseline architecture.
The main overhead comes from additional 16~KB SRAM required for each VRC.
Overall, the additional hardware introduces around 14\% area overhead (0.25~mm$^2$), compared to GSCore (1.78~mm$^2$) in 16nm.

\para{Software Setup.}
We evaluate on three large-scale datasets: Urban~\cite{lin2022capturing}, Mega~\cite{turki2022mega}, and HierGS~\cite{kerbl2024hierarchical}, as well as three small-scale datasets: T\&T~\cite{Knapitsch2017}, DB~\cite{hedman2018deep}, and M360~\cite{barron2022mipnerf360}.
To assess the effectiveness of \proj, we evaluate against three large-scale 3DGS algorithms: HierGS~\cite{kerbl2024hierarchical}, CityGS~\cite{liu2024citygaussian}, and OctreeGS~\cite{ren2024octree}.
The main difference of those algorithms is the LoD search.
For rendering quality, we adopt three widely used metrics: peak signal-to-noise ratio (PSNR), structural similarity index (SSIM), and perceptual similarity (LPIPS).
To mimic real VR scenarios, all stereo images are rendered at $2064\times2208$ resolution with a pupil baseline of 6~cm.

\para{Software Baselines.}
To evaluate the quality of our stereo rasterization, we compare against three baselines:
\begin{itemize}
    \item \mode{Base}: the baseline algorithm that renders both eyes.
    \item \mode{Warp}~\cite{chaurasia2020passthrough+}: a widely-used warping technique which uses a classic densification to fill disocclusions.
    \item \mode{Cicero}~\cite{feng2024cicero}: a state-of-the-art warping-based method, which uses neural rendering to fill disocclusions.
\end{itemize}
Note that, both \mode{Warp} and \mode{Cicero} use the generated depth map from 3DGS~\cite{chung2024depth} rather than the ground truth depth, which is not available in real-world scenarios.

\para{Hardware Baselines.}
To evaluate the efficiency of our hardware design, we compare three hardware baselines:
\begin{itemize}
    \item \mode{GPU}: a mobile Ampere GPU on Nvidia Orin SoC~\cite{orinsoc}.
    \item \mode{GSCore}~\cite{lee2024gscore}: a dedicated accelerator for 3DGS.
    \item \mode{GBU}~\cite{ye2025gaussian}: a hardware module that accelerates rasterization while the remaining operations are executed on the same mobile GPU as \mode{GPU}. For fairness, we implement 128 Row PEs in GBU to align with GSCore. 
    
\end{itemize}
We develop the RTL implementation of both \mode{GPU} and \mode{GSCore}.
The numbers of both RTL designs are then scaled down to 8~nm node to match Nvidia Orin~\cite{orinsoc}.

\para{Application Scenarios.} Two scenarios are compared to show the effectiveness of our collaborative rendering:
\begin{itemize}
    \item \underline{\textit{Video Streaming}}: a VR device communicates wirelessly with a remote server, which executes all the rendering operations and streams the compressed video streams via a standard H.265 video compression~\cite{h265}. The VR device is primarily used for display and lightweight processing, e.g., decoding and tracking.
    \item \underline{\textit{Collaborative Rendering}}: this scenario simulates a VR device that communicates wirelessly with a remote server.
    However, this scenario evaluates our collaborative rendering paradigm, where the server executes only the LoD search and streams the compressed Gaussians to the local device.
    The remaining rendering stages are processed on the local accelerator.
\end{itemize}
In both scenarios, the wireless communication energy is modeled as 100~nJ/B~\cite{liu2022augmented} with a data rate of 100~Mbps to model a high-speed Wi-Fi network~\cite{networkspec}. 
The remote server has two Nvidia A100 GPUs~\cite{a100} with 80 GB of memory to render left-eye and right-eye images separately.
\section{Evaluation}
\label{sec:eval}

\subsection{Rendering Quality}
\label{sec:eval:acc}

% \begin{figure}[t]
%     \centering
%     \includegraphics[width=\columnwidth]{stereo_psnr}
%     \caption{Rendering quality evaluation on stereo warping.
%     }
%     \label{fig:stereo_warp}
% \end{figure}

\begin{figure}[t]
\centering
\subfloat[PSNR evaluation.]{
	\label{fig:stereo_psnr}
    \includegraphics[width=\columnwidth]{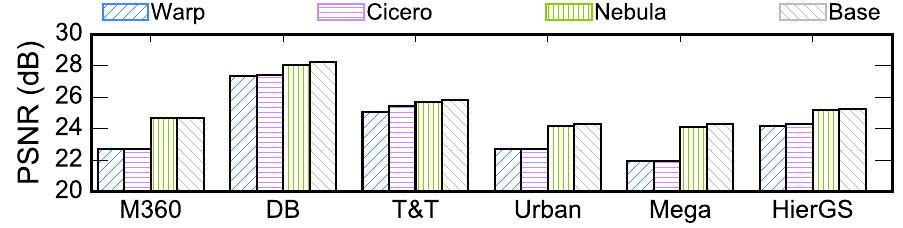}
    }
\\
\subfloat[SSIM evaluation.]{
	\label{fig:stereo_ssim}
	\includegraphics[width=\columnwidth]{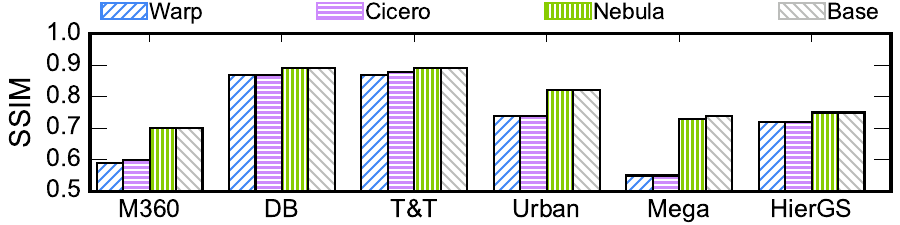}
}
\\
\subfloat[LPIPS evaluation.]{
	\label{fig:stereo_lpips}
	\includegraphics[width=\columnwidth]{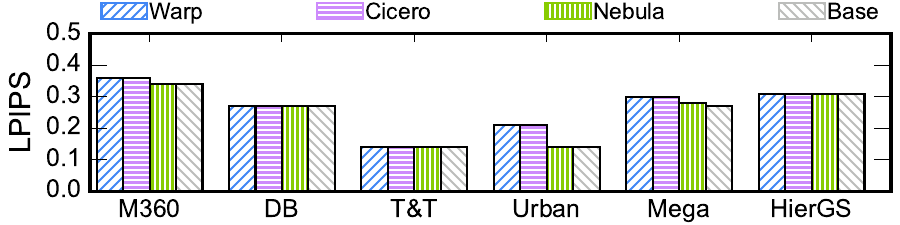}
} 
\caption{Rendering quality evaluation on stereo warping.
}
\label{fig:stereo_warp}
\end{figure}

\para{Stereo Rendering.}
We first evaluate the end-to-end visual quality of stereo rendering, as shown in \Fig{fig:stereo_warp}.
Here, \mode{Base} renders both eyes using HierGS~\cite{kerbl2024hierarchical}, while both \mode{Warp} and \mode{Cicero} generate the right-eye view by warping the left-eye image.
As expected, both \mode{Warp} and \mode{Cicero} introduce noticeable accuracy loss against \mode{Base}.
In contrast, \mode{\proj} delivers nearly identical quality to \mode{Base}, with only a 0.1~dB PSNR loss.
In the other two metrics, SSIM and LPIPS, \mode{\proj} shows no quality loss.
Meanwhile, this minor accuracy loss arises \textit{solely} from our compression scheme, since our stereo rasterization itself is bit-accurate.
% The results of SSIM and LPIPS are in \Sect{sec:appendix:acc}.

\begin{figure}[t]
    \centering
    \includegraphics[width=\columnwidth]{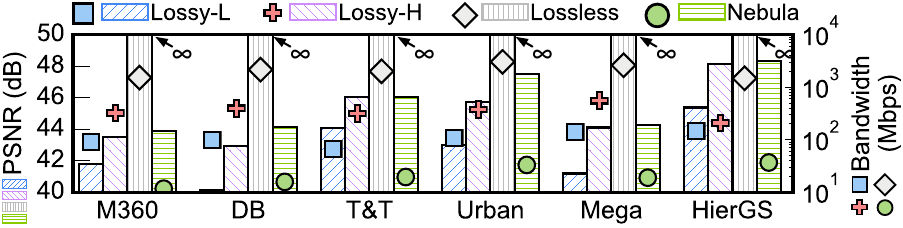}
    \caption{Rendering quality and bandwidth comparison on different compression methods..
    }
    \label{fig:compression_psnr}
\end{figure}

\para{Compression.}
\Fig{fig:compression_psnr} compares the visual quality of \proj with conventional video streaming using H.265 video compression~\cite{h265}.
We evaluate H.265 at three compression levels: \mode{Lossy-L}, \mode{Lossy-H}, and \mode{Lossless}.
L/H stands for low- and high-quality lossy compression.
% To highlight the effectiveness of our technique, 
We measure quality, PSNR, against the baseline rendering results instead of the dataset ground truth.
The results show that \proj preserves high quality, similar to \mode{Lossy-H}.
However, \proj can achieve significantly lower bandwidth under the 90~FPS VR resolution by exploiting temporal similarity.
Unless otherwise specified, we use \mode{Lossy-H} as the default compression scheme in the remaining evaluation.
% The common household Wi-Fi bandwidth is around 100~Mbps~\cite{networkspec}.

\subsection{Performance and Energy}
\label{sec:eval:perf}

\begin{figure}[t]
    \centering
    \includegraphics[width=\columnwidth]{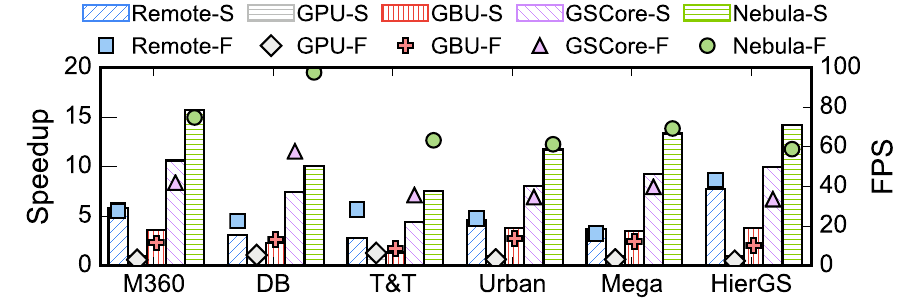}
    \caption{Overall performance comparison. All numbers are normalized to \mode{GPU}. ``S'': speedup; ``F'': frame per second.}
    \label{fig:overall_speedup}
\end{figure}

\begin{figure}[t]
    \centering
    \includegraphics[width=\columnwidth]{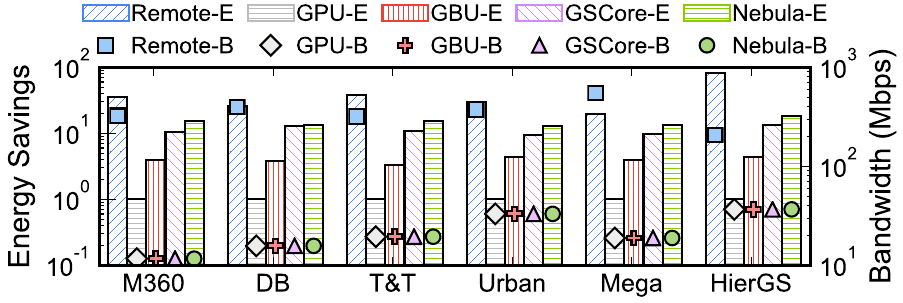}
    \caption{Overall energy and bandwidth savings. Display power is excluded since it is constant across methods. Numbers are normalized to \mode{GPU}. ``E'': energy; ``B'': bandwidth.}
    \label{fig:overall_energy}
\end{figure}

We first give the overall performance comparison by rendering two eyes, with $2064\times2208$ pixels per eye in VR.
% , we then show the performance gains from individual components.

\begin{figure}[t]
\centering
\begin{minipage}[t]{0.48\columnwidth}
  \centering
  \includegraphics[width=\columnwidth]{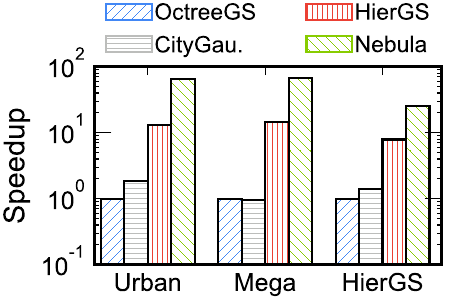}
  \caption{Speedup on LoD search. Temporal-aware LoD search achieves better performance than prior methods~\cite{ren2024octree, kerbl2024hierarchical, liu2024citygaussian}.}
  \label{fig:lod_speedup}
\end{minipage}
\hspace{2pt}
\begin{minipage}[t]{0.48\columnwidth}
  \centering
  \includegraphics[width=\columnwidth]{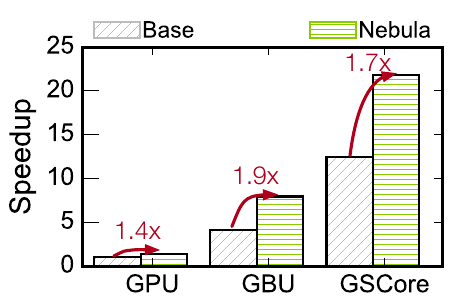}
  \caption{The speedup of \proj against the baseline algorithm on the client-side. Performance numbers are normalized to \mode{GPU}.}
  \label{fig:raster_speedup}
\end{minipage}
\end{figure}

\para{Speedup.}
\Fig{fig:overall_speedup} presents the overall performance comparison.
Here, \mode{Remote} applies the video streaming setup as in \Sect{sec:exp}, while others all adopt collaborative rendering proposed in this paper.
All variants execute HierGS~\cite{kerbl2024hierarchical} since it has the best LoD search performance in \Fig{fig:lod_speedup}.
We report motion-to-photon latency, normalized to the \mode{GPU} baseline.

Overall, \mode{\proj} achieves the highest speedup, 12.1$\times$, compared to \mode{GPU}.
In contrast, \mode{Remote} merely achieves 4.6$\times$, due to the network constraint.
Meanwhile, we also show the frame rate of different methods.
Here, we assume rendering and data communication can be pipelined, similar to prior work~\cite{xie2021q}.
Overall, \mode{\proj} achieve 70.1~FPS.
While \mode{\proj} does not achieve the VR requirement, 90~FPS, \Sect{sec:eval:sens} shows that \proj can easily achieve 90~FPS by scaling up VRC.

\para{Energy Savings.}
\Fig{fig:overall_energy} shows the overall energy savings against \mode{GPU}.
\mode{Remote} achieves the highest savings (38.4$\times$).
This is because when \mode{Remote} offloads all rendering computations to the remote GPU, the major energy consumption of the local device is simply wireless communication.
Among the collaborative rendering variants, \mode{\proj} delivers the best efficiency, achieving 1.4$\times$ and 14.9$\times$ lower energy compared to \mode{GSCore} and \mode{GPU}, respectively.

We also show the bandwidth requirement to achieve 90 FPS in \Fig{fig:overall_energy}, all collaborative rendering variants require less bandwidth (1925\%) than directly streaming videos.

\para{LoD Search.}
\Fig{fig:lod_speedup} shows the performance comparison of different algorithms on LoD search.
The original LoD search in OctreeGS~\cite{ren2024octree} is used as the baseline, and we compare HierGS~\cite{kerbl2024hierarchical} and CityGau~\cite{liu2024citygaussian}.
\mode{\proj} achieves much higher speedup (up to 52.7$\times$) than other methods by exploiting temporal similarity to eliminate redundant node accesses.

\para{Local Rendering.}
\Fig{fig:raster_speedup} shows the speedup of our stereo rasterization on local rendering (including preprocessing, sorting, and rasterization) over six datasets.
Across all architectural designs, \proj consistently delivers 1.4$\times$, 1.9$\times$, and 1.7$\times$ speedups on \mode{GPU}, \mode{GBU}, and \mode{GSCore}, respectively.

\begin{figure}[t]
\centering
\begin{minipage}[t]{0.48\columnwidth}
  \centering
  \includegraphics[width=\columnwidth]{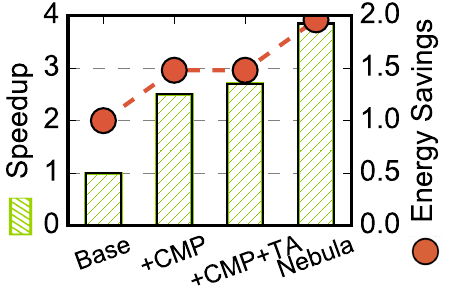}
  \caption{Ablation study on \proj. \mode{TA}: apply temporal-aware LoD search; \mode{CMP}: apply compression scheme; \mode{SR}: apply stereo rasterization.}
  \label{fig:ablation}
\end{minipage}
\hspace{2pt}
\begin{minipage}[t]{0.48\columnwidth}
  \centering
  \includegraphics[width=\columnwidth]{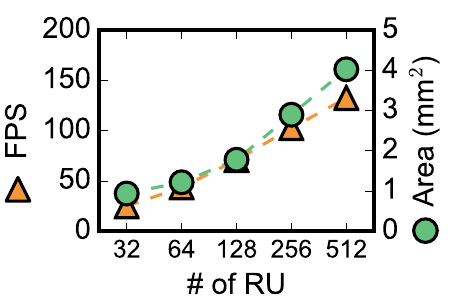}
  \caption{The scalability of performance and area to the number of rendering unit in VRC. \proj can easily achieve 90~FPS by scaling up.}
  \label{fig:ru_scalability}
\end{minipage}
\end{figure}

\subsection{Ablation Study}
\label{sec:eval:abl}

\Fig{fig:ablation} shows the ablation study of the contributions in \proj.
\mode{Base} executes the HierGS algorithm with collaborative rendering on our architecture.
We show the speedup (left y-axis) and energy (right y-axis) under: 
1) \mode{Base} with only our compression scheme (CMP), 
2) \mode{Base} with CMP and temporal-aware LoD search (TA), 
3) \mode{Base} with all optimizations.
On large-scale datasets, \mode{Base+CMP} achieves 2.5$\times$ speedup and 1.5$\times$ energy savings, \mode{Base+CMP+TA} achieves 2.7$\times$ speedup and 1.5$\times$ energy savings.
All together, \mode{\proj} achieves 3.9$\times$ speedup and 2.0$\times$ energy savings.

\subsection{Sensitivity Study}
\label{sec:eval:sens}

\para{RU Scalability.}
We demonstrate that \proj can readily meet VR frame rate requirements by scaling the rendering units (RUs) in the VRC. 
\Fig{fig:ru_scalability} shows the performance trend by averaging the results from three large-scale datasets~\cite{kerbl2024hierarchical, lin2022capturing, turki2022mega}. 
We show that doubling the RUs from 128 to 256 in the default VRC configuration enables our architecture to achieve real-time VR performance. 
However, increasing from 128 to 256 RUs increases the area by 62.9\%.

\begin{figure}[t]
\centering
\begin{minipage}[t]{0.48\columnwidth}
  \centering
  \includegraphics[width=\columnwidth]{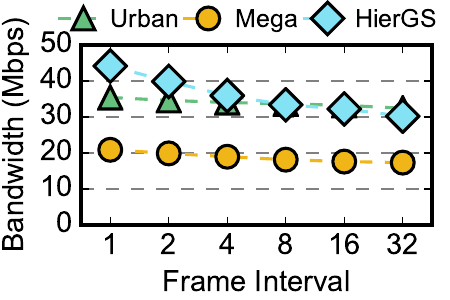}
  \caption{The sensitivity of bandwidth requirement to the frame interval, $w$.}
  \label{fig:frame_interval}
\end{minipage}
\hspace{2pt}
\begin{minipage}[t]{0.48\columnwidth}
  \centering
  \includegraphics[width=\columnwidth]{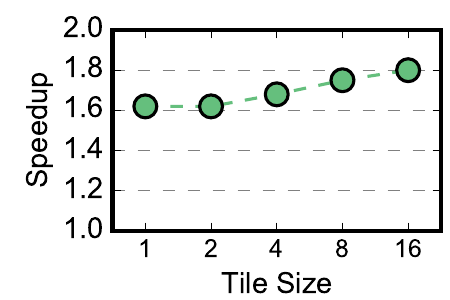}
  \caption{The sensitivity of performance to the tile size.}
  \label{fig:tile_size}
\end{minipage}
\end{figure}

\para{Frame Interval, $w$.}
By default, \proj performs temporal-aware LoD search once every four frames, which retains bandwidth requirements much lower than conventional video streaming. 
In \Fig{fig:frame_interval}, we further show that the bandwidth demand is largely insensitive to the choice of frame interval under three large-scale datasets. 
As the interval $w$ decreases, i.e., performing LoD search every $w$ frames, the bandwidth required for sustaining 90 FPS increases only modestly.

\para{Tile Size.}
Lastly, \Fig{fig:tile_size} shows the sensitivity of rendering performance to the tile size using HierHS dataset~\cite{kerbl2024hierarchical}.
The speedup is normalized to the corresponding baseline with the same tile size.
We observe that the performance decreases modestly as the tile size decreases.
This is because stereo rasterization helps mitigate warp divergence in rasterization.
As the tile size decreases, warp divergence diminishes.

\section{Related Work}
\label{sec:related}

\para{Collaborative Rendering.}
There are a few cloud-client collaborative rendering methods\mbox{~\cite{xie2021q, xu2023edge, he2020collabovr, ke2023collabvr, leng2019energy, feng2024cicero, zhao2020deja}} in literature. 
However, all current techniques are designed for mesh-based rasterization pipelines, not for 3DGS. 
For instance, Cicero\mbox{~\cite{feng2024cicero}} and CollabVR\mbox{~\cite{ke2023collabvr}} offload all rendering tasks to the cloud and only perform lightweight warping at the client side to accommodate disocclusions or stereo display. 
Both E-VR\mbox{~\cite{leng2019energy}} and DejaView\mbox{~\cite{zhao2020deja}} focus on 360 video streaming. 
Both leverage the unique features in 360 videos, e.g., the spatio-temporal redundancies or the user's field of view, to reduce the reprojection overhead and network traffic. 
Meanwhile, both Q-VR\mbox{~\cite{xie2021q}} and EDC\mbox{~\cite{xu2023edge}} incorporate foveated rendering to reduce the on-device workload.

However, all these methods continue to face bandwidth limitations when targeting higher FPS/resolution. 
More importantly, they partition the workload at the pixel level, which is incompatible with 3DGS since clients still require compute-intensive LoD search to filter Gaussians.
We propose \proj, a novel cloud-client co-design framework tailored for 3DGS and substantially reduce the on-device workload while minimizing the network traffic.

% Compared to video streaming methods above, this hybrid approach can reduce the amount of data transmitted over the network, as only partial rendering results or scene assets need to be sent~\cite{leng2019energy, meng2020coterie, bhojan2020cloudygame}.

\para{3DGS Acceleration.}
Recent studies propose various 3DGS architectures~\cite{feng2025lumina, ye2025gaussian, lee2024gscore, li2025uni, lee2025vr, lin2025metasapiens, durvasula2025arc, he2025gsarch, huang2026splatonic, li2023sltarch, zhang2025streaming}.
A few studies~\cite{lee2024gscore, ye2025gaussian, lin2025metasapiens, feng2025lumina, lee2025vr} are designed for the acceleration of the forward pass in 3DGS.
For instance, MetaSapiens~\cite{lin2025metasapiens} and GBU~\cite{ye2025gaussian} address the workload imbalance during rasterization. 
Lumina~\cite{feng2025lumina} proposes a caching technique to avoid redundant computation.
Some propose solutions for 3DGS training~\cite{he2025gsarch, durvasula2025arc}.
For example, ARC~\cite{durvasula2025arc} addresses the atomic operations in training while GSArch~\cite{he2025gsarch} prunes redundant gradient updates.
In principle, \proj is orthogonal to those works and can be applied to any existing 3DGS acceleration framework to support large-scale 3DGS rendering.

\para{Warping Techniques.}
Image warping~\cite{chaurasia2020passthrough+, szeliski2022image} is a lightweight technique for synthesizing novel views in conventional image-based rendering~\cite{chen2023view, chen1995quicktime}.
It exploits spatial and temporal correlations across frames to avoid redundant computations~\cite{buckler2018eva2, feng2020real, zhu2018euphrates, feng2023fast, ying2022exploiting, feng2019asv, zhao2020deja, zhao2021holoar}.
\proj also leverages both those similarities, but unlike prior work, \proj achieves bit-accurate rendering while delivering speedup.

\section{Conclusion}
\label{sec:conc}

Human imagination is boundless, and so too should be virtual 3D Gaussian worlds.
\proj marks the first step toward real-time, infinite-scale 3DGS splatting in VR.
By leveraging the vast resources of the cloud, we introduce a collaborative cloud–client rendering framework that alleviates communication bottlenecks and a novel stereo rasterization pipeline that eliminates redundancy in VR stereo rendering.

\begin{acks}

Thanks to \textit{Weikai Lin} from University of Rochester for the insightful conversation!
This work was supported by National Key R\&D Program of China under Grant 2022YFB4501400, the National Natural Science Foundation of China (NSFC) Grants (62532006 and 62402312), and Shanghai Qi Zhi Institute Innovation Program SQZ202316.

\end{acks}

\bibliographystyle{ACM-Reference-Format}
\bibliography{reference}

\end{document}